\documentclass[usenatbib]{mn2e}
\usepackage{times}
\usepackage{epsfig}
\usepackage{amsmath}
\title[]{Implications of the remarkable homogeneity of galaxy groups and clusters}
\author[Balogh \& McGee]{Michael L. Balogh$^{1}$, Sean L. McGee$^{1}$ 
\\
$^{1}$Department of Physics and Astronomy, University of Waterloo, Waterloo, Ontario, N2L 3G1, Canada\\
}
\date{\today}

\def\gtrsim{\mathrel{\raise0.35ex\hbox{$\scriptstyle >$}\kern-0.6em
\lower0.40ex\hbox{{$\scriptstyle \sim$}}}}
\def\lesssim{\mathrel{\raise0.35ex\hbox{$\scriptstyle <$}\kern-0.6em
\lower0.40ex\hbox{{$\scriptstyle \sim$}}}}

\def\k04{{^{0.4}K_s}}
\def\s04{{^{0.4}S1}}

\begin{document} 
\maketitle
\begin{abstract}
We measure the diversity of galaxy groups and clusters with mass $M>10^{13}h^{-1}M_\odot$, in terms of
the star formation history of their galaxy populations, for the purpose
of constraining the mass scale at which environmentally-important
processes play a role in galaxy evolution.
We consider three different group catalogues,
selected in different ways, with photometry and spectroscopy from the
Sloan Digital Sky Survey.  For each system we measure the fraction of
passively-evolving galaxies within $R_{200}$ and brighter than either
$M_r=-18$ (and with $z<0.05$) or $M_r=-20$ (and $z<0.1$).  We use the
$(u-g)$ and $(r-i)$ galaxy colours to distinguish between star-forming
and passively-evolving galaxies.  By considering
the binomial distribution expected from the observed number of members
in each cluster, we are able to either recover the intrinsic scatter in
this fraction, or put robust 95\% confidence upper-limits on its
value.  The intrinsic standard deviation in the fraction of passive
galaxies is consistent with a small value of $\lesssim 0.1$ in most
mass bins for all three samples. There is no strong trend with mass; even groups with
$M\sim10^{13}h^{-1}M_\odot$ are consistent with such a small, intrinsic
distribution.  
We compare these results with theoretical
models of the accretion history to show that, if environment plays a
role in transforming galaxies, such effects
must occur first at mass scales far below that of rich clusters, at most
$M\sim 10^{13}M_\odot$.
\end{abstract}
\begin{keywords}
galaxies: clusters
\end{keywords}

\section{Introduction}\label{sec-intro}
In the highly successful model of dark matter-dominated hierarchical galaxy formation,
clusters of galaxies grow over time by accreting matter from their
surroundings, with a well-defined distribution of halo masses ranging
from isolated galaxies to large groups
\citep[e.g.][]{LC94,ZMJB,Berr+08,McGee-accretion}.  If galaxy
evolution is sensitive to the mass of the host dark matter halo,
then differences in mass accretion history should be reflected in the
population residing within groups and clusters \citep{McGee-accretion}.  For example, the increase
in the fraction of young galaxies in cluster cores as a function of
redshift \citep[e.g.][]{BO84,Margo} can be linked to a corresponding increase in infall
rate \citep{Erica, KB01,Haines+09}, and the change in population with distance
from the cluster centre can be used to infer a transformation timescale
\citep[e.g.][]{infall}.  

Recently, \citet{McGee-accretion} have shown that the
cluster-to-cluster scatter in galaxy populations is a potentially
powerful indicator of ``pre-processing'' --- environmentally-driven transformation that may
have occurred in galaxies before they were accreted into the current
structure \citep[e.g][]{ZM98,Tornado-PP}.  For example, if the only
environmental effect on galaxies occurs at mass scales well below that
of rich clusters, one should expect a very homogeneous cluster population.
On the other hand, if galaxies are transformed only once they are
accreted into very massive systems, than the stochasticity of this
process will lead to more diversity in present-day cluster
populations.  In fact, such an argument was recently invoked by \citet{Pogg05},
who claim that an increase in diversity in groups with velocity dispersions below
$\sigma\sim500$km/s is evidence that transformations are likely
occurring at those scales.  However,
the uncertainties on the fraction of
galaxies with [OII] emission in a single system are large, and they increase with
decreasing mass.  Thus it is not clear whether or not the {\it
  intrinsic} scatter is such a strong function of mass.

With the advent of very large surveys of nearby galaxies, the
populations of galaxy groups and clusters and their correlations with
mass, X-ray properties and redshift are fairly well established
\citep[e.g.][]{F+01,deP_BO,PopessoI,W+05,Weinmann_1+06,A+07,Finn08,Kimm+09,Hansen+09,Bark+09,LGBB}.
However, relatively little attention has been given to the variation between
clusters.  One important exception is \citet{PBRB}, who analyzed 79
X-ray clusters in the Sloan Digital Sky Survey (SDSS), and find a
{\it r.m.s.} scatter of $0.19$ in the fraction of blue galaxies.  However, they do not
have a large enough sample to study this as a function of cluster mass,
which is crucial for identifying a putative transformation scale.
Furthermore, the measured {\it r.m.s.} includes a contribution from the
statistical uncertainties on individual measurements, and thus is an
upper limit on the intrinsic variation.  Very recently,
\citet{Haines+09} use {\it Spitzer} data of 30 X-ray luminous clusters at $z<0.3$ to
measure the fraction of strongly star-forming galaxies ($\gtrsim 8 M_\odot\mbox{yr}^{-1}$), and find a remarkably
small intrinsic scatter, consistent with zero once the trend with redshift is
accounted for.

In this paper, we revisit the issue using three different samples of
galaxy clusters, based on the SDSS.  Our results are consistent with
those of \citet{Pogg05} and \citet{PBRB}, but we take the extra step of
measuring the intrinsic cluster-to-cluster variation, and comparing
this with the model predictions of \citet{McGee-accretion}.  Throughout
the paper we adopt a cosmology with $\Omega_m=0.3$,
$\Omega_\Lambda=0.7$, and parameterize the Hubble constant as
$H_\circ=100h\mbox{km~s}^{-1}\mbox{Mpc}^{-1}$.

\section{Sample selection}\label{sec-data}
For our purposes, we require a large, homogeneous sample of galaxy
groups and clusters, together with a simple, reliable and sensitive
measurement of the galaxy population within them.  The SDSS \citep{SDSS_tech_short}, with its
large size, homogeneous data, and highly complete spectroscopy, is
particularly well-suited to this type of study.  We use data from the
DR6 spectroscopic sample, making use of
the NYU-VAGC of \citet{Blanton_cat}.  The data are unbiased for $r\leq
17.77$, and colours and luminosities are measured from Petrosian magnitudes,
k-corrected to $z=0.1$ using {\sc kcorrect} \citep{kcorrect}.  We make no
correction for spectroscopic completeness; this is a small correction
with little dependence on luminosity or colour \citep[e.g.][]{PBRB}, so has no impact on
these results.

There are many different ways to find galaxy clusters, usually based on
either galaxy position (with or without redshifts, and possibly using
colour information) or X-ray emission from the intracluster plasma.  The
ideal sample is complete, uncontaminated, and has a reliable, observable
property that can be related to the total mass of the system.  For our purposes,
completeness is perhaps the most critical, as this allows us to put
robust upper limits on the scatter we observe.  We will use three
different cluster catalogues for this analysis.  The first, which we
call the ``Halo'' catalogue, is taken from \citet{YMvdBJ}, who use some prior information
from theory and observation to associate every galaxy with a ``halo''.
The mass is then given by rank-ordering the groups by their total
luminosity or stellar mass (we use the latter), and associating that with a theoretical
dark matter mass function.  The second catalogue, also based on optical
data, is from \citet{Berlind+06}.  We refer to this as the ``FOF'' catalogue, as it is a
traditional friends-of-friends algorithm, with parameters calibrated to
match numerical simulations.  Masses for these systems are calculated
from their velocity dispersions, assuming virial equilibrium.
Finally, we consider an X-ray selected
sample of clusters, which consist of all 
HIFLUGCS \citep{HIFLUGCS} clusters in SDSS, and C4 \citep{C4} clusters
cross-correlated with the {\it ROSAT} catalogue.  For clusters in the
HIFLUGCS sample, we use the masses measured by \citet{HIFLUGCS}, under
the assumption of hydrostatic, isothermal gas; these have typical
uncertainties of 10--30 per cent.  For the C4 clusters, we
use the $M-L_x$ relation of \citet{HIFLUGCS} to convert {\it ROSAT}
luminosities to mass.  

Galaxy properties are known to correlate strongly with both
clustercentric radius, and the limiting luminosity or stellar mass of
the galaxy sample \citet[e.g.][]{PBRB}.  We will select all galaxies\footnote{Only
clusters with full SDSS coverage at $r<R_{200}$ are included in our analysis.} within the estimated
projected virial radius $r_{200}$, determined from the cluster
mass\footnote{We define $R_{200}$ to be the radius within which the
  total matter density is 200 times the critical density of the Universe.}.  This is done in different
ways for each sample.  The X-ray sample uses $r_{200}$ computed by
\citet{HIFLUGCS} for the HIFLUGCS clusters, and the results of
\citet{PBB}, based on fitting a King profile to the galaxy
distribution, for the {\it ROSAT}-matched C4 clusters.  For the FOF
group catalogue we calculate $r_{200}$ from the velocity dispersions,
using equation 8 in \citet{Finn05}.  The masses of the Halo clusters
from \citet{Y07} are measured by ranking the clusters by stellar mass
and comparing to dark matter mass functions, but the mass calculated is
for a radius with a {\it matter} overdensity of 180 which, for their
choice of cosmology, corresponds to 43 times the critical density.
Therefore their masses and radii are about 2.2 times larger than
$M_{200}$ and $R_{200}$, assuming a flat potential.  We  apply this
correction to make them more comparable to the other two samples.
However, we need not concern ourselves overmuch with ensuring complete
consistency between samples; indeed, differences between them help show
explicitly that our main conclusions are completely independent of
these details.
%, while for the Halo clusters we
%use $R_{180}$ following equation 5 in \citet{Y07}. The latter is
%typically 70\% larger than the former; rather than try to use a
%consistent calibration for all samples, we retain these differences to
%show explicitly the independence of our conclusions on such details.  
Both statistical and systematic uncertainties \citep[e.g. projection of
field galaxies along the line of sight,][]{PBRB} associated with this
measurement will only lead to increase the scatter we measure; thus our upper
limits will still be robust.  

We will select galaxies based on luminosity, and will consider
two versions of each sample.  The main catalogue is
limited to $z<0.05$, and we use all galaxies brighter than
$M_r+5\log{h}=-18$; for comparison we consider a ``bright'' sample
consisting of $M_r+5\log{h}<-20$ galaxies, and including clusters out to $z=0.1$.
Our results on the scatter are comparable for both samples.

\section{Definition of passive galaxies}
\begin{figure}
\leavevmode \epsfysize=8cm \epsfbox{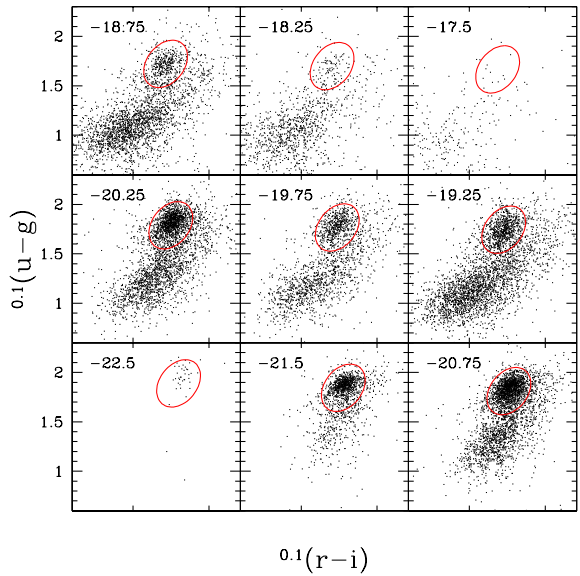} 
\caption{Colour-colour diagrams (k-corrected to z=0.1) for a random
  subset of SDSS galaxies in our sample within
  0.1 mag of the $M_r$ magnitude indicated in the top-left corner ($h=1$).  The
  galaxies are clearly separated into two populations, and we
  associate those galaxies within the red ellipses as
  ``passive'' galaxies, with little or no ongoing star formation.  The
  ellipse is fixed in size and orientation, but the centre is a linear
  function of $M_r$.  \label{fig-ccdiag} }
\end{figure}

We will take advantage of the well known fact that galaxies appear to
primarily divide into two classes.  One has predominantly red colours,
and little star formation, while the other consists of blue and actively
forming stars \citep[e.g.][and many others]{Strateva01_short,Baldry03,Kimm+09,Pozz+09}.  Most simply, this distinction can be made
using a colour-magnitude diagram to isolate red-sequence and blue-cloud
galaxies \citep[e.g.][]{Baldry03}.  However it is also known that a
single optical colour cannot
distinguish between dusty, star-forming galaxies, and truly old,
passive galaxies \citep[e.g.][]{PBRB}.  However, \citet{C17-dust,STAGES-dust} have shown that this distinction
can be reliably made using two colours: one bracketing the 4000\AA\
break, and another at longer wavelengths.  We show this for our full
sample of SDSS galaxies, in Figure~\ref{fig-ccdiag}; each panel shows
the $(u-g)$ colour as a function of $(r-i)$, for galaxies within $0.1$
magnitude of the $r-$band luminosity shown (all colours are k-corrected
to $z=0.1$).  The existence of two
populations is remarkably clear, as is the fact that a cut in $(u-g)$
colour alone would include galaxies from both populations.

We choose to select passive galaxies as those within the ellipse shown
in Figure~\ref{fig-ccdiag}.  The centre of the ellipse is a smooth
function of $M_r$, but the orientation and size is kept fixed.  These
ellipses are not optimized in any rigorous way, but reasonable
variations in definition have no influence on our results.  

For comparison we will also measure a simple ``red fraction'', based
on the $(u-r)$ colour as a function of $M_r$. We fit the red sequence
with a slope of $-0.0818$, and a colour $(u-r)=2.8$ at $M_r+5\log{h}=-20$, and
choose red galaxies to be those that are up to 0.25 mag bluer than this
line.  Our
results on the scatter of the red fraction are completely insensitive
to the details of this choice.

\section{Results}\label{sec-res}
\begin{figure}
\leavevmode \epsfysize=8cm \epsfbox{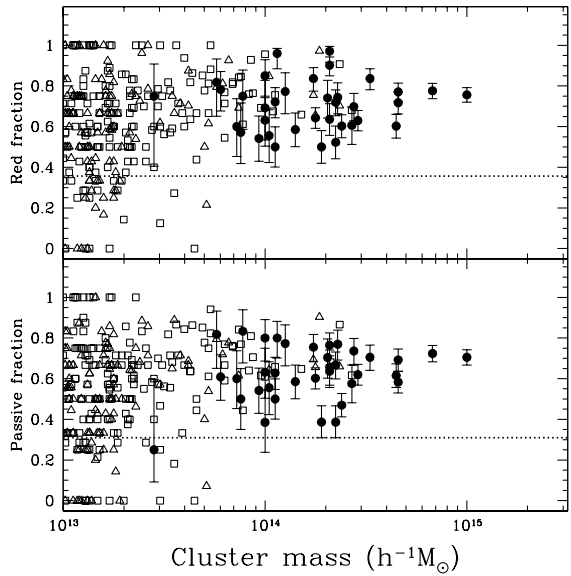} 
\caption{{\bf Bottom panel: } The fraction of passive galaxies in each
  cluster within the 
  optical Halo catalogue \citep[][open squares]{YMvdBJ}, the FOF
  catalogue \citep[][open triangles]{Berlind+06} and the X-ray
  matched catalogue (filled circles).   In all cases we use the main
  catalogue, with $M_r+5\log{h}<-18$.  Passive galaxies are defined
  as those within the elliptical areas in the colour-colour diagram of
  Figure~\ref{fig-ccdiag}.  The {\it dotted line} shows the passive
  fraction in the full SDSS sample (31\%), considering $M_r+5\log{h}<-18$ galaxies.
  Error bars are 68\% confidence limits using the binomial probability
  distribution, and are shown only on the X-ray sample, for clarity;
  the uncertainties on individual points at low masses get very large.  {\bf Top panel:} Similar, but here we show
  the fraction of {\it red} galaxies, defined using only the $(u-r)$
  colour.  \label{fig-red_fracs} }
\end{figure}

In Figure~\ref{fig-red_fracs} we show the fraction of passive galaxies
in every cluster in the three catalogues, as a function of
its mass, for the ``main'' sample ($M_r+5\log{h}<-18$). 
Uncertainties on the fractions are
computed using the full binomial distribution \citep{Gehrels}.
We see that both the Halo and X-ray samples of clusters are in good
agreement where they overlap, despite the different selection criteria,
and definitions of mass
and radius. 
The passive fractions are systematically larger than the $M_r+5\log{h}<-18$
sample as a whole (31\%, shown as the horizontal line), on average.
There is a weak trend with mass, such that the most massive clusters
have larger passive fractions, on average. The passive fractions are
systematically lower than the
{\it red} fractions based only on the $(u-r)$ colour (top panel), due to the contamination from
dusty-spiral galaxies in the latter.  However, the scatter from
cluster-to-cluster is similar whether we consider the red or
truly passive fraction.

The scatter in Figure~\ref{fig-red_fracs} becomes very large at lower
masses.  This is very similar to the trends 
shown in Figs. 4b and Fig.6 of \citet{Pogg05}, where they find the
scatter in the fraction of galaxies with [OII]-emission increases
sharply in systems with velocity dispersion
below $\sigma\sim 500$ km/s ($M\sim 1\times 10^{14}h^{-1}M_\odot$).
However, the uncertainties on individual
measurements are also much larger in these lower mass systems, since
they have fewer members.  Our task now is to recover the intrinsic
scatter from these observations.

\begin{figure}
\leavevmode \epsfysize=8cm \epsfbox{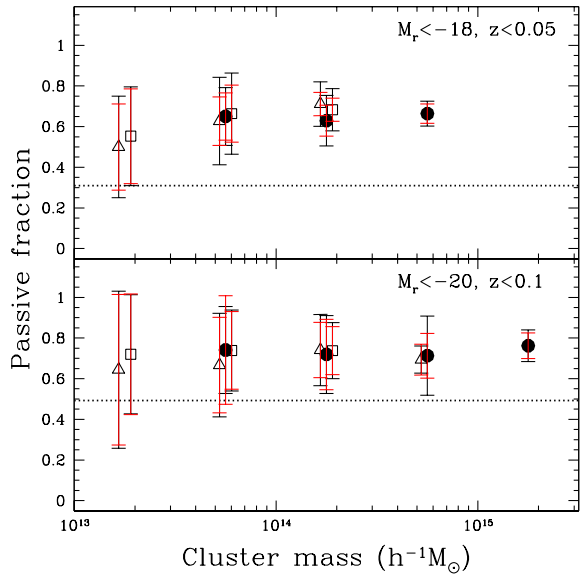} 
\caption{{\bf Top panel: }The points show the {\it average} passive fraction among all
  clusters, in bins of equal mass; only points with at least five
  contributing clusters are shown.  The
  {\it filled circles} represent the X-ray catalogue, while the
  {\it open squares} are the Halo catalogue of \citet{YMvdBJ} and the
  {\it open triangles} are the FOF catalogue of \citet{Berlind+06}. Black ``error
  bars'' indicate the standard deviation within each bin (not the error
  on the mean).  Red bars represent the standard deviation expected due
  to binomial sampling fluctuations alone, assuming no intrinsic
  scatter.  The horizontal, {\it dotted line} indicates the passive
  fraction in the whole $M_r<-18$ SDSS sample, 31\%.  The observed variance
  is typically very similar to that expected from the statistical
  uncertainties alone.  {\bf Bottom panel: }The same, for the $M_r+5\log{h}<-20$
  sample, which has more clusters but fewer galaxies per cluster.  Here
  the horizontal, {\it dotted line} indicates the passive
  fraction in the brighter $M_r<-20$ SDSS sample, 49\%.\label{fig-scatter} }
\end{figure}

In Figure~\ref{fig-scatter} we show the average passive fraction as a
function of mass, for both
samples.  The black ``error bars'' show the standard deviation in this 
fraction (they are not the error on the mean, which is much
smaller).  
The red ``error bars'' show the standard deviation expected from
statistical uncertainties alone, assuming {\it no} intrinsic
variation.  To compute this we treat the observed fraction for each
cluster as randomly drawn from a binomial probability distribution defined by
the number of galaxies in the cluster and with an expectation value given
by the mean value in each mass bin. We neglect any error in the mean
value itself, which is small.

This demonstrates that the variance in passive fraction is not
much larger than expected due to the statistical uncertainties
alone. The close agreement between all three samples also suggests that
systematic uncertainties related to sample selection, and mass or
radius definition, contribute much less to the observed variance than the
statistical uncertainties.

\begin{figure}
\leavevmode \epsfysize=8cm \epsfbox{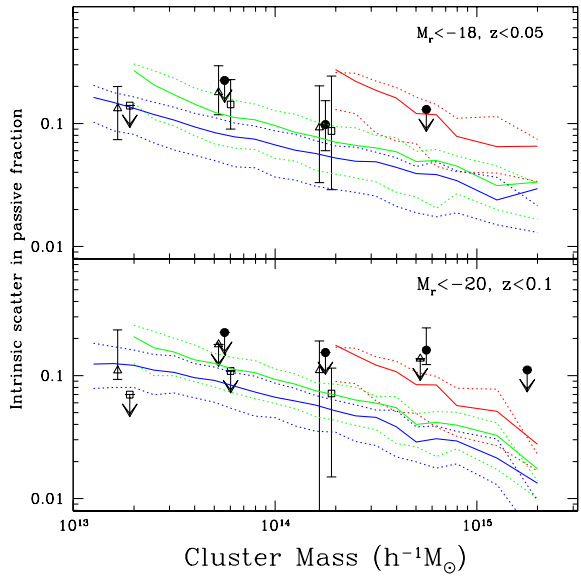} 
\caption{We show the intrinsic 1$\sigma$ {\it r.m.s} scatter on the passive fraction from
  cluster to cluster, in mass bins with at least five clusters.  The X-ray clusters are indicated with {\it filled
    circles}; the Halo clusters with {\it open squares}, and the FOF
  clusters with {\it open triangles}.  Where the
  observed variance is significantly larger than the statistical
  uncertainty, we have represented the 95\% confidence limits with
  error bars.  In cases where we are unable to
  measure a significant intrinsic scatter, in light of the statistical
  uncertainties, we have placed the point at the 95\% confidence upper
  limit, indicated with a downward arrow.  Model predictions from
  \citet{McGee-accretion} are shown as the lines.  Red, green and blue lines indicate
  $log(hM_{\rm trunc})=14$, 13 and 12, respectively.  In each case the
  solid line represents $T_{\rm trunc}=3$Gyr, with the dotted lines
  representing 1 Gyr (lower scatter) and 6 Gyr (higher scatter).  {\it Top panel:} The main
  cluster samples, used throughout most of this sample, restricted to
  $M_r+5\log{h}<-18$ and $z<0.05$.  {\it Bottom panel: } The larger, ``bright''
  cluster sample, restricted to $M_r+5\log{h}<-20$ and $z<0.1$.  
 \label{fig-scatter3} }
\end{figure}

For each mass bin we simulate the binomial distribution 10000 times,
and compute the variance of each simulation.  By adding in quadrature
an assumed amount of intrinsic (Gaussian) variation, ranging from 0 to
0.25, we estimate the 5\% and 95\% confidence limits on this
intrinsic scatter, and specifically the probability that it is
non-zero.  When the latter is greater than 95\%, we measure the
instrinsic scatter as the quadrature difference between the 
observed and predicted statistical standard deviations; otherwise, we
report an upper limit.
In
Figure~\ref{fig-scatter3} we show these measurements and upper
limits on the scatter, as a function of mass.  Generally, the scatter
is remarkably small, consistent with $0.1$ or less.  Importantly there is at most a
weak trend with cluster mass: even at $M\sim 10^{13}h^{-1}M_\odot$ the
intrinsic variation is small.  The fact that a large scatter is
observed in these systems is due to the fact that the uncertainties
associated with an individual group are large, limited by the number of
members.  With a large enough sample, however, it is possible to
overcome these uncertainties and allows us to put interesting limits on the
intrinsic scatter.

\section{Discussion and Conclusions}
We have shown that the fraction of passive galaxies in clusters has
remarkably small scatter between systems, generally 0.1 or less.  
We
can compare this directly with the 
predictions of \citet{McGee-accretion}, who use the galaxy formation
model of \citet{Font+08} to calculate the rate at which galaxies are
accreted into groups and clusters, and the amount of time they spend
within haloes of a given mass during their history.  More specifically, \citet{McGee-accretion}
show the fraction of
galaxies that have been inside a halo of mass $M>M_{\rm trunc}$ for a
time $t\geq T_{\rm trunc}$.  Under the assumption that such galaxies
are the only ones that turn passive, we can compare those predictions
with our results.  Although these predictions make use of
a semi-analytic model to trace galaxies, the results are determined
primarily by the dark matter growth history, which is taken from the
Millennium simulation \citep{Mill_sim}, and are largely undependent of
the recipes used to model the galaxies.  

We plot predictions from \citet{McGee-accretion} together with our
data, in Figure~\ref{fig-scatter3}.   Note that the observed scatter will be affected by
other systematic effects, such as field contamination and uncertainties
in the mass and radius; thus all of our points are best thought of us strict upper limits on
the true cluster population, which is what is being modeled.  Lines are shown for three
different $M_{\rm trunc}$ and three different $T_{\rm trunc}$, as
described in the caption.  It is interesting first of all that the
scatter predicted by the models is quite comparable to that observed,
even in the lowest mass groups; there is no evidence that the variance
in group properties significantly exceeds what would be expected from these simple
infall-based models.  The fact that these groups have
passive fractions significantly greater than the global average, with a small
intrinsic scatter, implies that $M_{\rm trunc}<10^{13}h^{-1}M_\odot$.  That
is, star formation in galaxies must be shutting down long before they
enter cluster-sized haloes.  Treating all the points as strict upper
limits, even the scatter in the most massive systems puts interesting
limits on these parameters.  If star formation is only truncated upon
accretion into clusters, $M_{\rm trunc}>10^{14}h^{-1}M_\odot$, then the
timescale must be $T\lesssim1$Gyr to be consistent with all the cluster
samples presented here.  Such a model, however, would predict greater
scatter than observed at $M=2\times10^{14}h^{-1}M_\odot$, and moreover that systems
with $M<M_{\rm trunc}$ should resemble the field population, which is
ruled out.  

The
results of \citet{Haines+09} potentially provide even stronger
constraints.  Their sample of clusters with masses of $\sim
2$--$20\times10^{14}M_\odot$ \citep{Okabe,Zhang+08} has an {\it rms} of 0.03
after accounting for a trend with redshift, and is consistent with no
intrinsic scatter.  At face value, this
suggests that not only must $M_{\rm trunc}$ be low, but $T_{\rm trunc}$
must also be small, $\lesssim 1$ Gyr.  However, their star formation
rate threshold is quite high, and this is probably why the average fraction of passive
galaxies at the lowest redshifts in their sample is $0.95$,
substantially higher than in the present sample.  This, together with
the wider redshift range covered, makes it difficult to directly
compare with our results.

In conclusion, the fraction of passive galaxies and the variance in
this fraction from system to system suggests that star formation is shut
off in galaxies within groups with masses 
$M\lesssim10^{13}h^{-1}M_\odot$; thus, ``pre-processing''
is crucial
to explain the observed properties of today's clusters.  These constraints can be significantly improved by
increasing survey depth, so there are more members per cluster, and by
increasing the volume surveyed so there are more contributing
clusters.  Mass measurements with smaller statistical uncertainty (for
example from X-ray observations with resolved temperature profiles) would
also be helpful in reducing the scatter, as there is a small trend for
the passive fraction to increase with mass.
Perhaps most importantly, repeating the analysis at higher redshift will be
valuable, as the scatter associated with large $T_{\rm trunc}$
models is predicted to increase significantly with redshift \citep{McGee-accretion}.

\section{Acknowledgments}\label{sec-akn}
The authors thank Chris Haines and the referee, Cristiano Da Rocha, for helpful comments that
significantly improved the paper.  
This research is supported by an NSERC Discovery grant to MLB, who
would also like to thank Bianca Poggianti, Richard Bower, David Gilbank and James Taylor for helpful
conversations about this work.
\bibliography{ms}

\begin{thebibliography}{45}
\expandafter\ifx\csname natexlab\endcsname\relax\def\natexlab#1{#1}\fi

\bibitem[{{Aguerri} {et~al.}(2007){Aguerri}, {S{\'a}nchez-Janssen}, \&
  {Mu{\~n}oz-Tu{\~n}{\'o}n}}]{A+07}
{Aguerri}, J.~A.~L., {S{\'a}nchez-Janssen}, R., \& {Mu{\~n}oz-Tu{\~n}{\'o}n},
  C. 2007, A\&A, 471, 17

\bibitem[{{Baldry} {et~al.}(2004){Baldry}, {Glazebrook}, {Brinkmann},
  {Ivezi\'{c}}, {Lupton}, {Nichol}, \& {Szalay}}]{Baldry03}
{Baldry}, I.~K., {Glazebrook}, K., {Brinkmann}, J., {Ivezi\'{c}}, Z., {Lupton},
  R.~H., {Nichol}, R.~C., \& {Szalay}, A.~S. 2004, ApJ, 600, 681

\bibitem[{{Balogh} {et~al.}(2000){Balogh}, {Navarro}, \& {Morris}}]{infall}
{Balogh}, M.~L., {Navarro}, J.~F., \& {Morris}, S.~L. 2000, ApJ, 540, 113

\bibitem[{{Barkhouse} {et~al.}(2009){Barkhouse}, {Yee}, \&
  {Lopez-Cruz}}]{Bark+09}
{Barkhouse}, W.~A., {Yee}, H.~K.~C., \& {Lopez-Cruz}, O. 2009, astro-ph/,
  0907.4800

\bibitem[{{Berlind} {et~al.}(2006)}]{Berlind+06}
{Berlind}, A.~A. {et~al.} 2006, ApJS, 167, 1

\bibitem[{{Berrier} {et~al.}(2009){Berrier}, {Stewart}, {Bullock}, {Purcell},
  {Barton}, \& {Wechsler}}]{Berr+08}
{Berrier}, J.~C., {Stewart}, K.~R., {Bullock}, J.~S., {Purcell}, C.~W.,
  {Barton}, E.~J., \& {Wechsler}, R.~H. 2009, ApJ, 690, 1292

\bibitem[{{Blanton} \& {Roweis}(2007)}]{kcorrect}
{Blanton}, M.~R. \& {Roweis}, S. 2007, AJ, 133, 734

\bibitem[{{Blanton} {et~al.}(2005)}]{Blanton_cat}
{Blanton}, M.~R. {et~al.} 2005, AJ, 129, 2562

\bibitem[{{Butcher} \& {Oemler}(1984)}]{BO84}
{Butcher}, H. \& {Oemler}, A. 1984, ApJ, 285, 426

\bibitem[{{De Propris} {et~al.}(2004)}]{deP_BO}
{De Propris}, R. {et~al.} 2004, MNRAS, 351, 125

\bibitem[{{Ellingson} {et~al.}(2001){Ellingson}, {Lin}, {Yee}, \&
  {Carlberg}}]{Erica}
{Ellingson}, E., {Lin}, H., {Yee}, H.~K.~C., \& {Carlberg}, R.~G. 2001, ApJ,
  547, 609

\bibitem[{{Fairley} {et~al.}(2002){Fairley}, {Jones}, {Wake}, {Collins},
  {Burke}, {Nichol}, \& {Romer}}]{F+01}
{Fairley}, B.~W., {Jones}, L.~R., {Wake}, D.~A., {Collins}, C.~A., {Burke},
  D.~J., {Nichol}, R.~C., \& {Romer}, A.~K. 2002, MNRAS, 330, 755

\bibitem[{{Finn} {et~al.}(2008){Finn}, {Balogh}, {Zaritsky}, {Miller}, \&
  {Nichol}}]{Finn08}
{Finn}, R.~A., {Balogh}, M.~L., {Zaritsky}, D., {Miller}, C.~J., \& {Nichol},
  R.~C. 2008, ApJ, 679, 279

\bibitem[{{Finn} {et~al.}(2005)}]{Finn05}
{Finn}, R.~A. {et~al.} 2005, ApJ, 630, 206

\bibitem[{{Font} {et~al.}(2008)}]{Font+08}
{Font}, A.~S. {et~al.} 2008, MNRAS, 389, 1619

\bibitem[{{Gehrels}(1986)}]{Gehrels}
{Gehrels}, N. 1986, ApJ, 303, 336

\bibitem[{{Haines} {et~al.}(2009)}]{Haines+09}
{Haines}, C.~P. {et~al.} 2009, ApJ, 704, 126

\bibitem[{{Hansen} {et~al.}(2009){Hansen}, {Sheldon}, {Wechsler}, \&
  {Koester}}]{Hansen+09}
{Hansen}, S.~M., {Sheldon}, E.~S., {Wechsler}, R.~H., \& {Koester}, B.~P. 2009,
  ApJ, 699, 1333

\bibitem[{{Kimm} {et~al.}(2009)}]{Kimm+09}
{Kimm}, T. {et~al.} 2009, MNRAS, 394, 1131

\bibitem[{{Kodama} \& {Bower}(2001)}]{KB01}
{Kodama}, T. \& {Bower}, R.~G. 2001, MNRAS, 321, 18

\bibitem[{{Lacey} \& {Cole}(1994)}]{LC94}
{Lacey}, C. \& {Cole}, S. 1994, MNRAS, 271, 676

\bibitem[{{Li} {et~al.}(2009){Li}, {Yee}, \& {Ellingson}}]{Tornado-PP}
{Li}, I.~H., {Yee}, H.~K.~C., \& {Ellingson}, E. 2009, ApJ, 698, 83

\bibitem[{{Lu} {et~al.}(2009){Lu}, {Gilbank}, {Balogh}, \& {Bognat}}]{LGBB}
{Lu}, T., {Gilbank}, D.~G., {Balogh}, M.~L., \& {Bognat}, A. 2009, MNRAS, 399,
  1858

\bibitem[{{Margoniner} {et~al.}(2001){Margoniner}, {de Carvalho}, {Gal}, \&
  {Djorgovski}}]{Margo}
{Margoniner}, V.~E., {de Carvalho}, R.~R., {Gal}, R.~R., \& {Djorgovski}, S.~G.
  2001, ApJL, 548, L143

\bibitem[{{McGee} {et~al.}(2009){McGee}, {Balogh}, {Bower}, {Font}, \&
  {McCarthy}}]{McGee-accretion}
{McGee}, S.~L., {Balogh}, M.~L., {Bower}, R.~G., {Font}, A.~S., \& {McCarthy},
  I.~G. 2009, MNRAS, 400, 937

\bibitem[{{Miller} {et~al.}(2005)}]{C4}
{Miller}, C.~M. {et~al.} 2005, AJ, 130, 968

\bibitem[{{Okabe} {et~al.}(2009){Okabe}, {Takada}, {Umetsu}, {Futamase}, \&
  {Smith}}]{Okabe}
{Okabe}, N., {Takada}, M., {Umetsu}, K., {Futamase}, T., \& {Smith}, G.~P.
  2009, ArXiv e-prints

\bibitem[{{Poggianti} {et~al.}(2006)}]{Pogg05}
{Poggianti}, B. {et~al.} 2006, ApJ, 642, 188

\bibitem[{{Popesso} {et~al.}(2007{\natexlab{a}}){Popesso}, {Biviano},
  {B{\"o}hringer}, \& {Romaniello}}]{PBB}
{Popesso}, P., {Biviano}, A., {B{\"o}hringer}, H., \& {Romaniello}, M.
  2007{\natexlab{a}}, A\&A, 464, 451

\bibitem[{{Popesso} {et~al.}(2007{\natexlab{b}}){Popesso}, {Biviano},
  {Romaniello}, \& {B{\"o}hringer}}]{PBRB}
{Popesso}, P., {Biviano}, A., {Romaniello}, M., \& {B{\"o}hringer}, H.
  2007{\natexlab{b}}, A\&A, 461, 411

\bibitem[{{Popesso} {et~al.}(2004){Popesso}, {B{\"o}hringer}, {Brinkmann},
  {Voges}, \& {York}}]{PopessoI}
{Popesso}, P., {B{\"o}hringer}, H., {Brinkmann}, J., {Voges}, W., \& {York},
  D.~G. 2004, A\&A, 423, 449

\bibitem[{{Pozzetti} {et~al.}(2009)}]{Pozz+09}
{Pozzetti}, L. {et~al.} 2009, ArXiv e-prints

\bibitem[{{Reiprich} \& {B{\" o}hringer}(2002)}]{HIFLUGCS}
{Reiprich}, T.~H. \& {B{\" o}hringer}, H. 2002, ApJ, 567, 716

\bibitem[{{Springel} {et~al.}(2005)}]{Mill_sim}
{Springel}, V. {et~al.} 2005, Nature, 435, 629

\bibitem[{{Strateva} {et~al.}(2001){Strateva}, {Ivezi{\' c}},
  {et~al.}}]{Strateva01_short}
{Strateva}, I., {Ivezi{\' c}}, {\v Z}., {et~al.} 2001, AJ, 122, 1861

\bibitem[{{Wake} {et~al.}(2005){Wake}, {Collins}, {Nichol}, {Jones}, \&
  {Burke}}]{W+05}
{Wake}, D.~A., {Collins}, C.~A., {Nichol}, R.~C., {Jones}, L.~R., \& {Burke},
  D.~J. 2005, ApJ, 627, 186

\bibitem[{{Weinmann} {et~al.}(2006){Weinmann}, {van den Bosch}, {Yang}, \&
  {Mo}}]{Weinmann_1+06}
{Weinmann}, S.~M., {van den Bosch}, F.~C., {Yang}, X., \& {Mo}, H.~J. 2006,
  MNRAS, 366, 2

\bibitem[{{Wolf} {et~al.}(2005){Wolf}, {Gray}, \& {Meisenheimer}}]{C17-dust}
{Wolf}, C., {Gray}, M.~E., \& {Meisenheimer}, K. 2005, A\&A, 443, 435

\bibitem[{{Wolf} {et~al.}(2009)}]{STAGES-dust}
{Wolf}, C. {et~al.} 2009, MNRAS, 393, 1302

\bibitem[{{Yang} {et~al.}(2005){Yang}, {Mo}, {van den Bosch}, \&
  {Jing}}]{YMvdBJ}
{Yang}, X., {Mo}, H.~J., {van den Bosch}, F.~C., \& {Jing}, Y.~P. 2005, MNRAS,
  356, 1293

\bibitem[{{Yang} {et~al.}(2007){Yang}, {Mo}, {van den Bosch}, {Pasquali}, {Li},
  \& {Barden}}]{Y07}
{Yang}, X., {Mo}, H.~J., {van den Bosch}, F.~C., {Pasquali}, A., {Li}, C., \&
  {Barden}, M. 2007, ApJ, 671, 153

\bibitem[{{York} {et~al.}(2000)}]{SDSS_tech_short}
{York}, D.~G. {et~al.} 2000, AJ, 120, 1579

\bibitem[{{Zabludoff} \& {Mulchaey}(1998)}]{ZM98}
{Zabludoff}, A.~I. \& {Mulchaey}, J.~S. 1998, ApJ, 496, 39

\bibitem[{{Zhang} {et~al.}(2008){Zhang}, {Finoguenov}, {B{\"o}hringer},
  {Kneib}, {Smith}, {Kneissl}, {Okabe}, \& {Dahle}}]{Zhang+08}
{Zhang}, Y., {Finoguenov}, A., {B{\"o}hringer}, H., {Kneib}, J., {Smith},
  G.~P., {Kneissl}, R., {Okabe}, N., \& {Dahle}, H. 2008, A\&A, 482, 451

\bibitem[{{Zhao} {et~al.}(2003){Zhao}, {Mo}, {Jing}, \& {B{\" o}rner}}]{ZMJB}
{Zhao}, D.~H., {Mo}, H.~J., {Jing}, Y.~P., \& {B{\" o}rner}, G. 2003, MNRAS,
  339, 12

\end{thebibliography}
\end{document}